\documentclass[10pt,superscriptaddress,amsmath,amssymb,aps,prl,twocolumn,showpacs]{revtex4}
\usepackage{mathrsfs}
\usepackage{graphicx}
\usepackage{subfigure}
\usepackage{dcolumn}
\usepackage{bm}
\usepackage{amssymb}
\usepackage{amsmath}
\usepackage{hyperref}
\usepackage{paralist}
\usepackage{color}

\newcommand{\tr}{\mathrm{tr}}

\def\be{\begin{equation}}
\def\ee{\end{equation}}
\def\bea{\begin{eqnarray}}
\def\eea{\end{eqnarray}}

\begin{document}

\title{Lower and upper bounds  of quantum battery power in multiple central spin  systems} 
\author{Li Peng}
\affiliation{State Key Laboratory of Magnetic Resonance and Atomic and Molecular Physics,
Wuhan Institute of Physics and Mathematics, 
Innovation Academy for Precision Measurement Science and Technology,  Chinese Academy of Sciences, Wuhan 430071, China}
\affiliation{University of Chinese Academy of Sciences, Beijing 100049, China.}

\author{Wen-Bin He}
\email[]{hewenbin18@csrc.ac.cn}
\affiliation{Beijing Computational Science Research Center, Beijing 100193, China}
\affiliation{The Abdus Salam International Center for Theoretical Physics, Strada Costiera 11, 34151 Trieste, Italy.}

\author{Stefano Chesi}
\affiliation{Beijing Computational Science Research Center, Beijing 100193, China}
\affiliation{Department of Physics, Beijing Normal University, Beijing 100875, China}
\affiliation{The Abdus Salam International Center for Theoretical Physics, Strada Costiera 11, 34151 Trieste, Italy.}

\author{Hai-Qing Lin }
\affiliation{Beijing Computational Science Research Center, Beijing 100193, China}
\affiliation{Department of Physics, Beijing Normal University, Beijing 100875, China}

\author{Xi-Wen Guan}
\email[]{xwe105@wipm.ac.cn}
\affiliation{State Key Laboratory of Magnetic Resonance and Atomic and Molecular Physics,
Wuhan Institute of Physics and Mathematics, 
Innovation Academy for Precision Measurement Science and Technology,  Chinese Academy of Sciences, Wuhan 430071, China}
\affiliation{NSFC-SPTP Peng Huanwu Center for Fundamental Theory, Xian 710127, China}
\affiliation{Department of Theoretical Physics, Research School of Physics and Engineering,
Australian National University, Canberra ACT 0200, Australia}

\begin{abstract}
We study the energy transfer process in  quantum battery systems consisting  of  multiple central spins and bath spins.
Here with ``quantum battery"  we refer to the central spins, whereas  the bath  serves as  the ``charger".
For the single central-spin battery, we analytically derive  the time evolutions  of the energy transfer and the charging power with arbitrary number of bath spins.
For the case of multiple central spins in the battery, we find  the scaling-law relation between the maximum power $P_{max}$ and the number of central spins $N_B$. 
It approximately satisfies  a  scaling  law relation $P_{max}\propto N_{B}^{\alpha}$, where scaling exponent $\alpha$ varies with the bath spin number $N$ 
from the lower bound $\alpha =1/2$ to the upper bound $\alpha =3/2$.
The lower and upper  bounds  correspond to the limits $N\to 1$ and   $N\gg N_B$, respectively. 
In thermodynamic limit, by applying the Holstein-Primakoff (H-P) transformation, we rigorously prove  that the upper bound is 
$P_{max}=0.72 B A  \sqrt{N} N_{B}^{3/2}$, which shows the same advantage in scaling of a recent charging protocol based on the Tavis-Cummins model.
Here $B$  and $A $ are the external magnetic field and coupling constant between the battery and the charger. 

\end{abstract}
\date{\today}
\maketitle

\section{I. Introduction} 

Energy resources are always an  important subject of modern sciences \cite{iea}, dating back to the fuel-coal energy to nuclear energy \cite{Gamow}, to present renewable energy including wind and solar energy \cite{iea,Dolf,Joel}.  
The exploitation of energy resources significantly involve  the study of the energy transfer, storage and  generation. 
Recently,  it attracts enormous attention to study quantum  heat engine \cite{Medley2011,  Chen2019} and refrigeration \cite{Weld2010, Yu2020, Peng2019, Wolf2011},  energy storage and transfer in quantum mechanical systems.  The latter are named as  ``quantum battery" \cite{Alicki, Hovhannisyan, Campaioli, Ferraro, Gian18, Campaioli2018, le2018spin, Lewenstein, Rossini2019, Andolina2019, Caravelli2020, Gian, campo, Sergi}. 
Classical electrical battery stores energy by electric field,  which can be understood in the frame of electrodynamics. 
In contrast, the quantum battery usually refers to the devices  that utilize the quantum degrees of freedom  to store and transfer energy. 
In general, the quantum degrees of freedom and their interplay can endow the quantum battery with advantage beyond classical picture.
In the last few years, there have been variety of methods to study the quantum battery, including  realization schemes, battery power and charging process \cite{Pirmoradian2019, yyzhang, An2020,  Rossini, Santos2019, Santos2020, Friis2018}.
In these studies, quantum coherence and entanglement seemed to play a key role in manipulation of quantum batteries.
R. Alicki and M. Fannes \cite{Alicki}  showed that entanglement can help extract more work in charging process. 
However, the role of entanglement in work extraction is still in  debate \cite{Hovhannisyan,Campaioli}.
D. Ferraro et.al \cite{Ferraro} showed that quantum advantage of charging power is manifested by an array of $N$ collective two-level systems in a cavity in comparison to  the  $N$ parallel  quantum battery cells of the  Dicke model.
G. M. Andolina et.al \cite{Gian18} considered the role of correlations in  different systems serving as a quantum battery, including the combination of two-level systems and quantum harmonic oscillators.
 There are  also other schemes to realize  quantum batteries, for example, using the  open systems \cite{Gherardini2020, Carrega2020, Farina2019, Barra,Watanabe}  and external field driving systems \cite{Crescente2020}. 

However, there  still remain many open  questions concerning  quantum battery. 
These mainly concern the battery's largest energy, power, extractable energy etc. 
Firstly, the number of quantum battery cells  can not be increased to infinity in order  to reach an infinite power. 
Therefore it imposes a theoretical and practical  challenge  to manipulate as many quantum battery cells as possible due to the decoherent nature of quantum systems. 
%
 Secondly, the number of quantum degrees of freedom in chargers is  usually not big  enough such that the transferred energy is not able to  saturate the full cells of a battery during the  charging process. 
 Nevertheless, both the numbers of quantum degrees of freedom and coupling strength between the battery and charger can alter the quality and power of the quantum battery. 
 This essentially involves the issue how the storage capability of quantum battery depends  on the cell numbers of both battery and charger.

In this paper, we study the energy transfer process in   quantum batteries of the multiple-central spin model.
Here the battery consists of $N_B$  spins which are displayed  in collective mode during the  charging process, whereas  the charger has $N$ bath spins, see the Fig.\ref{fig_spinc}.
We analyze  the dependence of the energy transfer  and the power of the battery  on the number of battery spins $N_B$ and  the number of the charger  spins  $N$. 
 We find  that the transferred  energy linearly increases with the number of battery spins  $N_B$ when $N$ and $N_B$ are comparable, then saturate to a certain value. 
 While  the maximum power monotonically  increases  with respect to the number  $N_B$  in a power-law form $P_{max} \propto N_{B}^{\alpha}$, where $\alpha$ shows a  dependence on the  number of charge spins  $N$.
 For the limit $N\ll N_B$, the lower bound reads $\alpha =1/2$. 
  For the case $N\gg N_B$, the  maximum  energy of battery always linearly increases with the number of   battery spins. 
  While for $N\gg N_B$ and in thermodynamic limit, the power-law relation of  the  maximum power $P_{max} \to N_{B}^{1.5}$ is verified  by  numerical calculation.
  In thermodynamic limit, using  the Holstein-Primakoff transformation, we also rigorously prove  that 
  $P_{max}=0.72 B \cdot A \sqrt{N} N_{B}^{\alpha}$, where the  exponent gives  the upper bound  $ \alpha=3/2$. 
  However, for $N_B$ incoherent batteries with single spins, we prove  that the maximum power is given by $P_{ max} \approx 0.72B A \sqrt{N} N_B$.
  Here $B$  and $A $ are  respectively the external magnetic field and coupling constant between the battery and the charger.
  It turns out the battery power essentially depends on the  cell numbers  of   the battery and the charger.
  %
  %
  Our analytical results shed light on the high-power charging  of quantum  batteries.

\section{II. The quantum battery and the model}
\begin{figure}
\begin{center}
\includegraphics[scale=0.35]{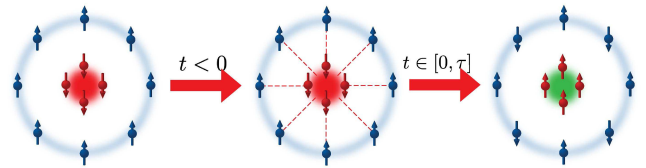}
\end{center} 
\caption{The illustration of charging of multiple central spin model working as quantum battery, whereas the bath spins serves as the charger. At $t<0$, there is no interaction between battery and charger. While interaction is switched on during the charging process  $t\in [0,\tau]$, the battery is charged. }
\label{fig_spinc}
\end{figure}

\emph{Quantum battery.}---In this section, we discuss  the basic setup of the quantum battery. 
The protocol  of the underlying quantum battery  consists  of two parts, i.e., the quantum reservoir of energy-battery $H_{B}$ and the energy charger $H_{C}$. 
Both  the battery and charger are composed of quantum particles  that  have  discrete energy levels  and degeneracies. 
The charging process is accomplished by switching on the interaction $H_I$ between the  battery and the charger so as  to complete the energy transfer, see Fig.~\ref{fig_spinc}.
For this purpose, the whole Hamiltonian of this model is given by 
\begin{equation}
H(t)=H_{B}+H_{C}+\lambda(t)H_{I},
\label{qb}
\end{equation}
where coupling constant $\lambda(t)$ will be used to control the charging period.
 It equals to 1 for one charging period $t \in [0,\tau]$ and is 0 for other  time. 
There exists   energy input and output between the battery and charger during the charging period  from  $t=0$ to  $t=\tau$. 
 The energy transfer, the charging speed and the power of the battery essentially depend on the number of the battery and charger, interaction strength between them and other external drives if possible.  

In order to comply with the  terminology which is used in the previous work \cite{Gian18,Ferraro},  we first  introduce the definitions of energy and power of the quantum battery. 
We consider that system which evolves unitarily  such that the wave function $\psi(t)$ describe the state of system. Meanwhile, the state of battery spins can be described by reduced density matrix of battery $\rho_B(t)=\tr_{C}[\vert \psi(t) \rangle \langle \psi(t) \vert]$, here $\tr_{C}$ denotes the  trace taking  over the spins in the charger.
The energy of the battery is defined as the expectation value of the Hamiltonian $H_B$
\begin{eqnarray}\label{eq_Ebc}
 	 E_B(t)&=&\tr[H_B\rho_B(t)].
\end{eqnarray}
Here $\rho_B$ denotes the reduced density matrix of the battery. 
The transferred  energy of quantum battery is given by $\Delta E_B(t)=E_B(t)-E_B(0)$, where $E_B(0)$ is the energy before charging process. 
 Meanwhile the charging power of the battery is defined as 
 \begin{equation}
 P_{B}(t)=\Delta E_B(t)/{t}.
 \end{equation}
Since the unitary evolution of the whole system during the charging period, the energy will flow between charger and battery back and forth. 
It is not necessary to track the energy  and power at every moment. 
Usually, one  chooses  the maximum energy  as a measure of the capability for storing energy $E_{max}=\max[\Delta E_B(t)]$,  and accordingly the maximum  power reads $P_{max}=\max[ P_{B}(t)]$. 

It has been demonstrated  \cite{Ferraro} that collective battery cells of two-level systems coupled to a cavity mode can enhance the energy transfer by manipulating the detuning between the two-level systems and the cavity mode. 
They argued that the collective evolution proceeds through states characterized by quantum entanglement among the battery cells.
In general, we naturally expect  an  existence of such quantum advantage generated during the  time evolution of the whole  many-particle systems of the Hamiltonian (\ref{qb}).  
Here  we aim to investigate   the scaling laws of the maximum energy  $E_{max}$ and the maximum power $P_{max}$  with respect to the numbers of battery spins.  
Similarly, in our work, the multiple central spins are prepared in a collective way, so that there also exists a certain form of quantum advantage in the system considered below.  
 Such scaling laws reveal coherent nature between the battery and the charger, as well as the quantum entanglement among the spin qubits  in the battery induced by the unitary evolution.   

\emph{The model.}---
In order  to realize a high-power quantum battery, we consider the multiple central spin model with the  Hamiltonian  (\ref{qb})  given by 
\begin{eqnarray}
H_{B}&=&B\mathbf{S} ^{z}, \label{H-B}\\
H_{C}&=&h \mathbf{J}^{z},\label{H_C} \\
H_I&=&A(\mathbf{S} ^{+} \mathbf{J}^{-}+\mathbf{S} ^{-} \mathbf{J}^{+})+2\Delta\mathbf{S} ^{z} \mathbf{J}^{z}. \label{H-I}
\label{Hhcs}
\end{eqnarray}
Here, for our  convenience, we denoted  the large  spin operators  $\mathbf{{S} ^{\alpha}}=\sum_{i=1}^{N_B} s_{i}^{\alpha},\alpha=\{z,+,-\}$, and $\mathbf{J}=\sum_{j=1}^{N}\mathbf{\tau}_{j}^\alpha $ for the battery and charger, respectively. 
We adopt different notations for central spins $s_{i}^{\alpha}$ and bath spins $\tau_{i}^{\alpha}$ in order to avoid a misunderstanding.
They  are both the  spin-$\frac{1}{2}$ operators. 
We regard the central spins as the storage cells of the quantum battery, while  the bath spins as charging energy carrier.
The energy can be exchanged between the battery and the charger  through spin-exchange interaction term $H_I$, also see the  Fig.\ref{fig_spinc}.
The $H_I$ contains the spin flip-flop interaction and the  Ising type interaction, which are respectively  denoted by $A$ and $\Delta$, i.e.   the exchange coupling constant and anisotropic parameter.
We also set the coupling strength $A=1$ for our rescaled units in the whole paper, see \footnote{For the unit of other parameters, we compared them with the $A$ to obtain their unit. At present, superconductor qubits may serve as quantum battery platform to observe the results of this work since spin exchange interaction can be realized experimentally. In practical experiment, spin-exchange coupling  usually takes the  unit $[{\mathrm time}]^{-1}$,  for instance in \cite{Guo:2021}, they set Hamiltonian as $H/\hbar$ and the spin-exchange coupling $J_{m,m+1}\sim 1/60 ns^{-1}$. }
The parameters $B$ and $h$ are the effective external magnetic fields for the central spins and bath spins, respectively. 
And $N_B$ is the number of central spins, $N$ is the number of bath spins. 

We introduce the Dicke state $\vert n \rangle= \vert \frac{N}{2},n- \frac{N}{2}\rangle $, which is the eigenstate of $\mathbf{J^{2}}$ and $\mathbf{J^{z}}$.
The Dicke state can be expressed as 
\begin{equation}
\vert n \rangle=\frac{1}{\sqrt{C^{n}_{N}}} \sum_{j_{1}<\cdots <j_{n}} \vert j_{1},\cdots ,j_{n} \rangle,
\end{equation}
here $\vert j_{1},\cdots ,j_{n} \rangle=\tau_{j_{1}}^{+}\cdots \tau_{j_{n}}^{+} \vert \Downarrow\rangle$, and normalization coefficient $C^{n}_{N}$ is combination number $\frac{N!}{n!(N-n)!}$,  and $ \vert \Downarrow\rangle$ denotes the down spins as the reference state.
The Dicke state  is highly entangled many-body quantum state. 
The action of the above spin  operators on state $\vert n \rangle$ are given by 
\begin{eqnarray*}
 \mathbf{J^{z}}\vert n \rangle &=& (-\frac{N}{2}+n)\vert n \rangle,\\
\mathbf{J^{-}}  \vert n \rangle &=&  \sqrt{b_{N,n}} \vert n-1 \rangle,\\
\mathbf{J^{+}} \vert n \rangle &=& \sqrt{b_{N,n+1}} \vert n+1 \rangle,
\end{eqnarray*}
where denoted the coefficient 
$b_{N,n}=n(N-n+1)$. 
For the large spin  operator of the battery $\mathbf{S}$, they have similar properties through  replacing $N$ by  $N_B$ and replacing the spin operators $\tau^\alpha_j$ by $s^\alpha _j$, respectively.  
We consequently  introduce the state basis of the whole system $\vert m,n \rangle$ for the degree of the battery $m\in \{0,1,\cdots,N_B\}$ and the degree of the charger $n\in \{0,1,\cdots,N\}$.
The Hamiltonian of the whole system $H$ can be diagonalized by the recurrence relation developed in \cite{He-WB:2019}. 
 For special case $N_B=1$, we can  analytically obtain the whole  dynamical evolution of spin polarization, see  the Appendix. 

\section{III. Numerical and analytical Results}
We first consider the numerical study of the general form of the quantum battery (\ref{qb}). 
We assume  the  initial state as 
\begin{align}
\vert  \Phi_{0}\rangle={\vert  \varphi_{0}\rangle}_{B} \otimes {\vert  \phi_{0}\rangle}_{C}. 
\end{align}
Usually,  the battery spins are in lowest states while the charger  is in the higher  excited states. 
For performing our numerical study, we choose the initial state as $\vert  \Phi_{0}\rangle=\vert 0, N \rangle=\vert \Downarrow, \Uparrow \rangle$.  
The wave function of system evolves with time, namely, 
\begin{equation}
\vert\psi(t)\rangle=\exp(-iHt)\vert  \Phi_{0}\rangle.
\end{equation}
By definition Eq. (\ref{eq_Ebc}), we may calculate the evolution of  the energy of battery as function of time $t$.

\subsection{A. Special $N_B=1$ case}
At the beginning of this subsection, we first study  the results of the special case $N_B=1$ with $\vert  \Phi_{0}\rangle=\vert \downarrow \rangle_{B} \otimes {\vert  \phi_{0}\rangle}_{C}$ in order  to get intuitive recognition of the  energy transfer. 
Usually one can choose the states of bath spins  as the Fock state or  spin coherent state. 
Here we consider the Fock state for  the initial state of the bath spins
\begin{align}
\vert  \Phi_{0}\rangle=\vert \downarrow \rangle \otimes {\vert  n\rangle},
\end{align}
where the bath spin state ${\vert n\rangle}$ represents  $n$ flipped spins among the $N$ spins. 
The time evolution of the wave function can be obtained from  the Hamiltonian $H$ with the Eqs. (\ref{H-B}-\ref{H-I}), i.e. 
\begin{equation}\label{wave_function}
\vert \psi(t)\rangle = e^{-i\theta t}\Big[P_{\uparrow}^{n}(t) \vert \uparrow \rangle \vert n-1\rangle+ P_{\downarrow}^{n} (t) \vert \downarrow \rangle \vert n\rangle \Big]. 
\end{equation}
Here the global phase $\theta$ can be omitted and the two probability amplitudes are given by $P_{\uparrow}^{n}=-i\frac{2\sqrt{b_{N,n}}A}{ \Omega_{n}}\sin( \frac{\Omega_{n}t}{2})$ and $
	P_{\downarrow}^{n}= i\frac{\Delta_{n}}{ \Omega_{n}}\sin( \frac{\Omega_{n}t}{2})+\cos( \frac{\Omega_{n}t}{2})
$. 
The wave function satisfies the normalization condition $|P_{\uparrow}^{n}|^2+|P_{\downarrow}^{n}|^2=1$. 
In the above equations, we denoted the parameters 
\begin{eqnarray}
\Delta_{n}&=&B-h+(2n-1-N)\Delta,\nonumber \\
 \Omega_{n}&=&\sqrt{\Delta_{n}^2+4b_{N,n}A^2}.\nonumber
 \end{eqnarray}

 Using the wave function (\ref{wave_function}), the charging energy and the power of quantum battery are  obtained explicitly  
\begin{small}
\begin{eqnarray}
\Delta E_{B}(t)& =& B\frac{4b_{N,n}A^2}{ \Omega_{n}^2} \sin^{2}( \frac{\Omega_{n}t}{2})
\label{Energy_BAI_3}\\
P_B(t)&=&\Delta E_{B}(t)/t= B\frac{4b_{N,n}A^2}{ \Omega_{n}^2\,t} \sin^{2}( \frac{\Omega_{n}t}{2}). \label{power_BAI_3}
\end{eqnarray}
\end{small}
The detailed  calculation can be found in Appendix, also see the calculation for the Jaynes-Cummings (JC) model \cite{Gian18}. 
Based on this result,  we briefly present a discussion on the energy transfer of the quantum battery below.

\begin{figure}
\begin{center}
\includegraphics[scale=0.6]{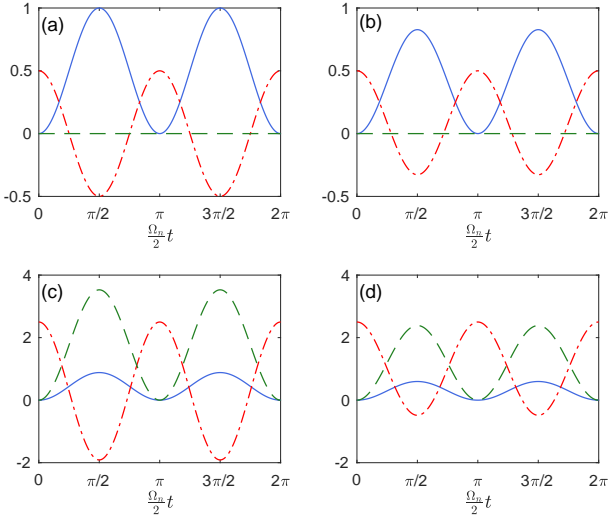}
\end{center} 
\caption{The charging energy of quantum battery $\Delta E_B(t)$(blue solid line), the energy of charger $\Delta E_C(t)$(dashed red line) and the interaction energy $E_I(t)$(dashed green line) are shown as function of $\Omega_{n}t/2$. 
(a) Charger and quantum battery are at resonant for $B=h=1$ and $\Delta=0$. 
(b) Charger and quantum battery are at resonant for $B=h=1$ and $\Delta=5$. 
(c) Charger and quantum battery are off from the resonance  for $B=5$, $h=1$ and $\Delta=0$. 
(d) Charger and quantum battery are off from  resonance  for $B=5$, $h=1$ and $\Delta=5$.
In  the above subplots(a)-(d), we set $A=1$, $N=10$ and $n=N/2=5$.
In the (a) and (b), the interaction energy are always equal to zero for  the whole time regime and the energy can be totally transferred from charger to battery. 
}
\label{diff_initialstate}
\end{figure}

(i) Resonant case $B=h$, $\Delta=0$,  the charging energy is   given by 
\begin{eqnarray}
	\Delta E_{B}(t)&=&B\sin^2(\sqrt{b_{N,n}}A t).
\end{eqnarray}
After an approximation, the maximum of the  power is given by
\begin{eqnarray}
	P_{max}\approx 0.72B A\sqrt{b_{N,n}}.\label{single-central-s-P}
\end{eqnarray}
From the expression of $\Delta E_B(t)$, the maximum transferred energy and the  consuming  time are given by 
\begin{equation}
	E_{max}=B ,\ \tau_{min}=\frac{\pi}{2A\sqrt{b_{N,n}}}
	\label{tmin}
\end{equation}
From the definition of $b_{N,n}$, we may obtain the minimum time to transfer the maximum energy, namely,   $\tau_{min}=\frac{\pi}{2A}\frac{1}{{(N+1)}/2}$, here we see  $n=\frac{N+1}{2}$. 
This means that the quantum battery is able to  store the maximum energy in the shortest time for the initial state with $n=(N+1)/2$  flipped bath spins.

(ii) Non-resonant case $B\neq h$ or $\Delta\neq 0$. In this case, we observe that the charging energy of quantum battery  $|\Delta E_B(t)/B | <1$ and interaction energy $E_I(t)=\langle H_{I}\rangle \neq 0$. 

In Fig.\ref{diff_initialstate} (a) (b), we show the results of the battery and charge at  the resonance. 
There is no interaction energy between the battery and charger in the figure (a). 
For the Fig.~\ref{diff_initialstate} (b),  we chose  $B=h$, $\Delta=5$, $n=N/2=5$, the terms involving the factors  $(N/2-n)$ and $(B-h)$  in the charging energy  of quantum battery vanish  (see Appendix Eqs. (A7) and (A8)).
In this case, the maximum energy intake is limited by $N$ due to the conservation of the energy.
Fig.~\ref{diff_initialstate} (c) (d) present the non-resonant case, at  which there exists an  interaction energy between the battery and charger. 
This indicates that the transferred energy from the  charger to the quantum battery is essentially subject to  the interaction form. 
 In this scenario, the maximum transfer energy strongly depends on $\Delta, \, B$ and $h$.

%

%
%
%

\subsection{B. Arbitrary $N_B$ case }
For arbitrary number of battery spins $N_B$ case, the eigenfunction is constructed by $\varphi=\sum_{m}\sum_{n} c_{m,n}\vert m, n \rangle$. 
After substituting the above ansatz  into eigenvalue equation, the superposition coefficient $c_{m,n}$ are determined by following recurrence equation
\begin{eqnarray}
w_{mn}c_{m,n}+A\sqrt{b_{N_B,m}b_{N,n+1}}c_{m-1,n+1}+\nonumber \\
A\sqrt{b_{N_B,m+1}b_{N,n}}c_{m+1,n-1}=E c_{m,n},\label{Recurrence}
\end{eqnarray}
where coefficient $b_{N_B,m}=m(N_B-m+1)$ is defined as $b_{N,n}$ previously, and $w_{mn}=B(-\frac{N_B}{2}+m)+h(-\frac{N}{2}+n)+2\Delta(-\frac{N_B}{2}+m)(-\frac{N}{2}+n)$. 
Here, for  the battery,  $m\in \{0,1,\cdots,N_B\}$ and for the charger $n\in \{0,1,\cdots,N\}$.
However, the recurrence equation    (\ref{Recurrence}) with two variables $m, n$ is  very difficult to be solved analytically. 
In order to  study the energy transfer, we exactly diagonalize the Hamiltonian to obtain the time evolution of the system.
Without losing the essential properties of the battery,  we   consider the interaction energy between charger and battery as zero  by choosing the parameter  $\Delta=0$ and $B=h=1$ in our numerical calculation. 
We will show that  for this case the Hamiltonian can map to the Tavis-Cummings model \cite{sm,JC,dicke}.
In addition, the system is prepared in the  initial state $\vert  \Phi_{0}\rangle=\vert \Downarrow, \Uparrow \rangle$, i.e. $m=0$ and $n=N$.  
The time evolution of the energy and power of the battery can be obtained numerically and analytically.

\begin{figure}
\begin{center}
\includegraphics[scale=0.6]{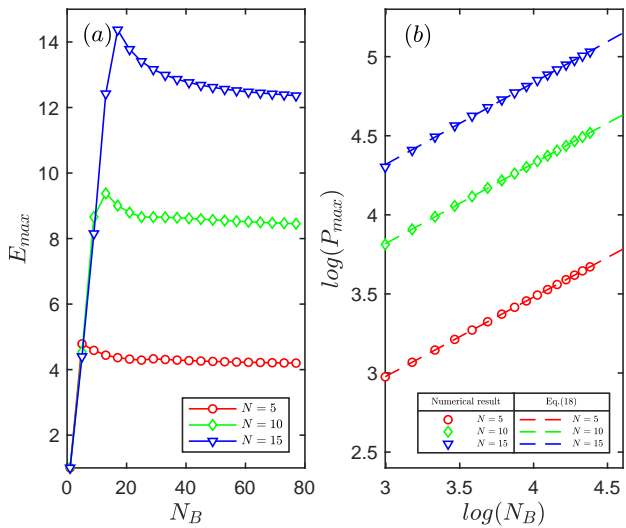}
\end{center} 
\caption{The maximum energy (a) and power (b) of the  multiple  central spin model v.s.  the number of battery spins $N_B$ for  different charger settings $N$. The dashed lines in (b) show  the numerical fitting of the power relation  Eq. (\ref{P_alpha})  in logarithmic scale  for $N_B \in [20,80]$, i.e. 
$N=5$,  $\alpha=0.5013$, $\beta=4.3706$ (red line);
$N=10$,  $\alpha=0.5067$,  $\beta=9.9668$ (green line)  and 
$N=15$,  $\alpha=0.5172$, $\beta=15.9241$ (blue line), which agree with the numerical results showing in the corresponding symbols. 
This confirms   the  lower bound of the scaling exponent of the maximum power  $\alpha \to 1/2$.
 Here we set  $A=1$, $B=h=1$, $\Delta=0$ with the initial state $n=N$, and $m=0$.}
\label{lmgEP}
\end{figure}

For a classical battery device, the electric current is static so that a charging process  can be complete in a certain  time. 
However,  for the quantum battery, the energy transfer is essentially  subject to dynamical evolution and  depends  not only on the devices but also on the charging time. 
Let's first understand  how the charging process depend on the number of the  battery spins when  the  number of charger spins is fixed. 
 If the battery spins are  token  as the Fock state like  that for the  charger spins,  the  dynamical evolution of the battery involves the highly entangled Dicke state $\vert m \rangle, m=1,...N_B$ in charging process.  
 Such kind of setting leads to a collective charging of the multiple central spin quantum battery, similar to the two-level system coupled to the single cavity  mode, i.e. the Dicke model  \cite{Ferraro}. 
By using Holstein-Primakoff transformation, we will  prove that our model can be mapped to Tavis-Cummings model, see the Eq.(\ref{Htc}) in analytical study part.  
Meanwhile, Tavis-Cummings model relates to the Dicke model  by the rotating wave approximation, see  \cite{Ferraro}.
Therefore, we naturally expect an existence of  a general scaling relation between  the battery power and the number of battery spins $N_B$ in the quantum battery of the Tavis-Cummings-like model. 
 After performing numerical calculation, we find that  the maximum power 
\begin{equation}
P_{max}\propto  \beta (N) N_{B}^{\alpha},
\label{P_alpha}
\end{equation}
where the exponent $\alpha$ is strongly affected by the number of charger spins and the initial state, $\beta $ is a function of the number of the charger spins  $N$. 
%
 %
Here  the scaling exponent $\alpha$  essentially marked a collective nature of  the battery in transferring  energy.

\begin{figure}
\begin{center}
\includegraphics[scale=0.6]{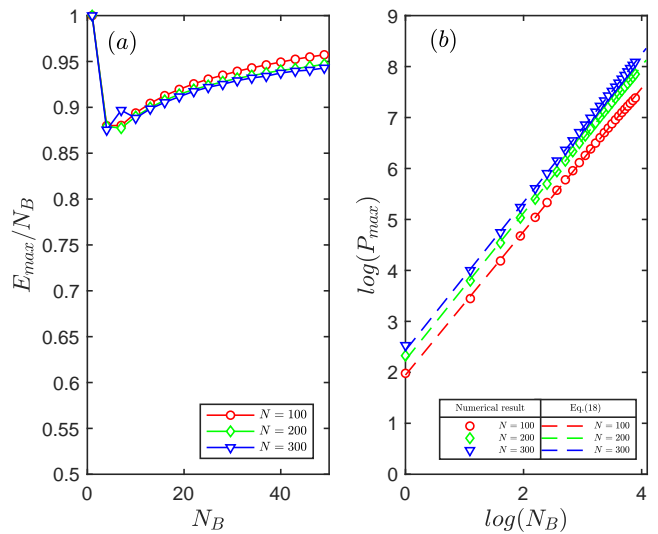}
\end{center} 
\caption{
The rescaled  maximum energy (a)  and the maximum  power  (b) v.s. the number of the  battery spins $N_B$ for different number of charger spins $N$. 
The dashed lines in (b) show  the numerical fitting of the power relation  Eq. (\ref{P_alpha})  in logarithmic scale  for $N_B \in [1,50]$, i.e. 
$N=100$,  $\alpha=1.4075$, $\beta=7.0056$ (red line);
$N=200$,  $\alpha=1.4434$,  $\beta=9.4058$ (green line)  and 
$N=300$,  $\alpha=1.4540$, $\beta=11.3456$ (blue line), which agree with the numerical results showing in the corresponding symbols. 
This agreement confirms   the  upper  bound of the scaling exponent of the maximum power  $\alpha \to 3/2$ in thermodynamic limit.
Here we set  $A=1$, $B=h=1$, $\Delta=0$ with the initial state $n=N_B$, and $m=0$.
}
\label{lmgEP_N}
\end{figure}

\begin{figure}[]
\begin{center}
\includegraphics[scale=0.4]{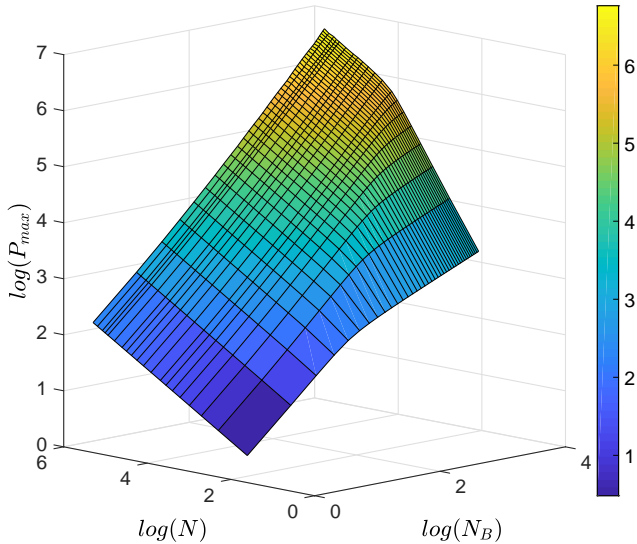}
\end{center}
\caption{Logarithmic contour plot of the maximum power v.s. the numbers of the battery spins $N_B$ and charger spins $N$. 
It clearly shows different values of  power scaling exponent $\alpha$ in the regimes $N_B\gg N$ and $N\gg N_B$. 
Here we set  $A=1$, $B=h=1$, $\Delta=0$ with the initial state $n=N$, and $m=0$.
}
\label{qbalpha_Nall}
\end{figure}



 Using the above setting and the initial state, i.e. $m=0,\, n=N$, we firstly compute the time evolution of energy and the maximum power,
more detailed explanation on the numerical calculation is given in Appendix.
%
%
In the Fig.\ref{lmgEP}, we show the maximum energy and maximum power as function of the number $N_B$ of battery spins for  different numbers of charger spins.
In Fig.~\ref{lmgEP} (a), we observe that the maximum energy $E_{max}$ increases linearly with respect to the number of battery spins $N_B$  when  $N_B<N$ and saturates  to a constant value when $N_B>N$. 
The  maximum energy clearly shows a kink.
%
%
In Fig.~\ref{lmgEP} (b), we observe  that the maximum power $P_{max}$ increases monotonically with respect to  the battery spins $N_B$  for  different number of charger spins $N=5$ (red circle),  $N=10$ (green square) and $N=15$ (blue triangle).
The logarithmic plot of the maximum power $P_{max}$ directly gives the  scaling exponent $\alpha$  which fits the relation (\ref{P_alpha}) for the region $N_B>N$, see Fig.~\ref{lmgEP}(b) and  the Appendix.
This result confirms  the  lower bound of the scaling exponent of the maximum power, i.e. $\alpha \to 1/2$, in the region $N_B>N$. 

In Fig.~\ref{lmgEP_N}, we  demonstrate the maximum energy and the power law relation (\ref{P_alpha}) of the  battery maximum power  for    $N\gg N_B$ with the initial condition $n=N_B$.
We observe that the rescaled  maximum energy $E_{max}/N_B$ does exhibit  plateaux in thermodynamic limit, see Fig.~\ref{lmgEP_N} (a).
In Fig.~\ref{lmgEP_N} (b), the plot of the maximum power $P_{max}$  in logarithmic scale show the scaling relation (\ref{P_alpha}) in agreement with the analytical result given in Eq. (\ref{pmaxhp}), where the analytical result 
$N=100$,  $\alpha=1.5$, $\beta=7.2$;
$N=200$,  $\alpha=1.5$,  $\beta=10.1823$  and 
$N=300$,  $\alpha=1.5$, $\beta=12.4708$.
Both the $N_B$ and $N$ take the thermodynamic limit, the result Eq. (\ref{pmaxhp}) can exactly hold.

In Fig.~\ref{qbalpha_Nall}, we further demonstrate the power law relation (\ref{P_alpha}) of the  maximum power with respect to  the numbers of battery spins $N_B$  and charger spins $N$, where we set  the initial state $n=N$ and  consider the ranges $N_B\in  [1,40]$ and $N\in  [1,200]$. 
This figure also confirms  the observation shown in Fig.~\ref{lmgEP} and Fig.~\ref{lmgEP_N}. 
Our numerical results  show that the collective battery is enable to enhance the power through increasing  the number of battery  cells when the charger resources are big enough.  In certain regions  there exist  lower and upper bounds of  the scaling exponents in the maximum power.
In next subsection, we will present an analytical  proof of  these two bounds.


\subsection{Analytical study }

In order to get a comprehensive understanding of the lower and upper bounds of the scaling exponent found  by numerics  in last section, we now present a rigorous calculation of the maximum energy and power of the quantum battery of Tavis-Cummings  type. 
If we apply the Holstein-Primakoff transformation to both the bath and battery spins, thus the  whole  Hamiltonian of system (\ref{H-B}-\ref{H-I}) can be mapped to the Tavis-Cummings model \cite{sm,JC,dicke}, where the $N_B$ central spins  are regarded as the  $N_B$ atoms of two-levels energy. 
 For  $N\gg1$, $N_B\gg1$, we apply transformation for charger spins
\begin{eqnarray}\label{HP-1}
	\mathbf{J}^{+}&=&\sqrt{N}a^{\dagger}\sqrt{1-a^{\dagger}a/N}\nonumber\\
	\mathbf{J}^{-}&=&\sqrt{N}\sqrt{1-a^{\dagger}a/N}a\nonumber\\
	\mathbf{J}^{z}&=&-\frac{N}{2}+a^{\dagger}a.
\end{eqnarray}
Without losing generality, we can obtain the Tavis-Cummings model for the case  $\Delta=0$
\begin{eqnarray}
H_{TC}&=& B\mathbf{S} ^{z}+h(a^{\dagger}a-\frac{N}{2})+A\sqrt{N}(\mathbf{S} ^{+} {a}+ \nonumber \\
&& \mathbf{S} ^{-}{a}^{\dagger}). 
\label{Htc}
\end{eqnarray}
And we continue to apply the Holstein-Primakoff  transformation to battery spins
\begin{eqnarray}\label{HP-2}
	\mathbf{S}^{+}&=&\sqrt{N_B}b^{\dagger}\sqrt{1-b^{\dagger}b/N_B}\nonumber\\
	\mathbf{S}^{-}&=&\sqrt{N_B}\sqrt{1-b^{\dagger}b/N_B}b\nonumber\\
	\mathbf{S}^{z}&=&-\frac{N_B}{2}+b^{\dagger}b.
\end{eqnarray}
In above formulas, $a(b)$ and $a^{\dagger}(b^{\dagger})$ both are the annihilation and creation  operators of boson. 
Substituting Eq.(\ref{HP-1}) and Eq.(\ref{HP-2}) into the Hamiltonian Eq. (\ref{H-B}-\ref{H-I}),  we can obtain
\begin{eqnarray}\label{TC-Hamiltonian}
	H&\approx& B(-\frac{N_B}{2}+b^{\dagger}b)+h(-\frac{N}{2}+a^{\dagger}a)\nonumber\\
	&& +A\sqrt{N_BN}(a^{\dagger}b+ab^{\dagger}).
\end{eqnarray}
Here we neglected the  terms  $a^{\dagger}a/N$ and $b^{\dagger}b/N_B$  since $N\gg1$, $N_B\gg1$,  
while we set $\Delta=0$ in the $H_I$ for simplifying our analytical study.
%
%
Later, based on the whole Hamiltonian (\ref{TC-Hamiltonian}),  we will analytically  derive the scaling laws of the maximum energy and the maximum power with respect to   the numbers of battery and charger spins. 
In this model, the total particle number is conserved and thus we have  $[H, a^{\dagger}a+b^{\dagger}b]=0$.
Without losing a generality,  we can choose the Hamiltonian as the following form for $B=h$
\begin{eqnarray}
	H_I=A\sqrt{N_B N}(a^{\dagger}b+ab^{\dagger}).
\end{eqnarray}
We take the initial state as previous $\vert  \Phi_{0}\rangle=\vert m,n\rangle=\vert m \rangle_{B} \otimes {\vert  n\rangle}_{C}$ and the quantum battery is in the lowest state, namely,  $m\rightarrow 0$. 
The maximum charging energy of the quantum battery is influenced by not only the energy levels of the battery and  charger but also  the choice of their initial states.
In  quantum optics, the energy levels of photons can be infinite.
 For the multiple central spins, the maximum transferred energy $\Delta E_B\propto B\cdot N_B$. 
 We reasonably choose $n-m \sim  N_B$, i.e. the charger contains enough energy to charge  the battery  to a level  of the maximum energy.
The wave function at time $t$  is given as the previous expression $ \vert\psi(t)\rangle=\exp(-\mathrm{i} H_{I}t)\vert  \Phi_{0}\rangle$.
By definition, the charging energy of the quantum battery is given by 
\begin{eqnarray}
	\Delta E_B(t)=B\Big[\left\langle \psi(t)\right|b^{\dagger}b\left|\psi(t)\right>-\left\langle \Phi_{0}\right|b^{\dagger}b\left|\Phi_{0}\right>\Big],
\end{eqnarray}

Let's further  define the  operator  
 \begin{eqnarray}\label{F_definition}
 	\hat{\mathbf{F}}=b^{\dagger}b-a^{\dagger}a.
 \end{eqnarray}
 Its time evolution  is given by
  \begin{eqnarray}
 	F(t)&=&\left\langle \Phi_{0}\right|e^{iH_I t} \hat{\mathbf{F}}e^{-iH_I t}\left|\Phi_{0}\right>. \label{Ft_definition}
 \end{eqnarray}
 
 After carefully calculating the recurrent  commutation relations between the operators $H_I$ and  $\hat{\mathbf{F}}$, 
 we obtain the following expression 
  \begin{eqnarray}\label{Ft_opeartor}
 	&e^{iH_I t}& \hat{\mathbf{F}}e^{-iH_I t}=\hat{\mathbf{F}}+\sum_n \frac{1}{n!}
 	[iH_I t,[iH_It,\cdots,[iH_I t,\hat{\mathbf{F}}]\cdots]]\nonumber\\
 	&=&\sum_{m=0}^{\infty}\frac{i^{2m+1}}{(2m+1)!}(2tA{\sqrt{N_BN}})^{2m+1}(a^{\dagger}b-ab^{\dagger})\nonumber\\
	&&+\sum_{m=0}^{\infty}\frac{i^{2m}}{(2m)!}(2tA{\sqrt{N_BN}})^{2m}\hat{\mathbf{F}}\nonumber\\
 	&=&i\sin(2A{\sqrt{N_BN}}t)(a^{\dagger}b-ab^{\dagger})\nonumber\\
	&&+\cos(2A{\sqrt{N_BN}}t)\hat{\mathbf{F}}.
 \end{eqnarray}
 Substituting Eq.~(\ref{Ft_opeartor}) and Eq.~(\ref{F_definition}) into Eq.~(\ref{Ft_definition}), we further obtain a simple expression 
 \begin{equation}
 	F(t)=(m-n)\cos(2A{\sqrt{N_BN}}t).
 \end{equation}
Moreover, the total particle number  $\hat{\mathbf{N}}=b^{\dagger}b+a^{\dagger}a$  is   a conserved quantity, i.e. $[H_I,\hat{\mathbf{N}}]=0$.
 Therefore we have $N(t)=\left\langle \psi(t)\right|\hat{\mathbf{N}}\left|\psi(t)\right>=m+n$. 
 It follows that 
 \begin{eqnarray}
 	\left\langle \psi(t)\right|b^{\dagger}b\left|\psi(t)\right>&=&\frac{N(t)+F(t)}{2}\nonumber\\
	&&=\frac{m+n}{2}+\frac{m-n}{2}\cos{(2A{\sqrt{N_BN}}t)}.\nonumber
 \end{eqnarray}
Thus the charging energy and the power  of the quantum battery are given by 
\begin{eqnarray}
	\Delta E_B(t)&=&B\cdot(n-m)\sin^2(A{\sqrt{N_BN}}t), \label{Ebhp} \\
	P_B(t)&=&B\cdot(n-m)\frac{\sin^2(A{\sqrt{N_BN}}t)}{t}  \label{pbhp},
\end{eqnarray}
respectively. 

 It is straightforward  to obtain the maximum power that is given by  ${P}_{max}=B\cdot 0.72A\sqrt{N_BN}(n-m)$  for a time $\tau=1.16/(A\sqrt{N_BN})$.
  As being mentioned in previous section, we demand $n-m \sim N_B$ and $N\gg N_B$,  thus the maximum power is  given by 
\begin{eqnarray}
{P}_{max}=0.72BA\sqrt{N}N_{B}^{3/2}
\label{pmaxhp}
\end{eqnarray}
that  reveals a significant advantage of this charging protocol, which leads to the upper bound of the scaling exponent $\alpha=3/2$. 
We observe that  in the early charging process,
the power reaches the maximum while the energy does not reach the maximum. 
This means that, the maximum power $P_{max} \propto N_{B}^{3/2}$ can indeed occur in the early time of the charging process, when the flipped spin $\langle b^{\dagger}b \rangle$ in the battery is much less than the number of battery cells $N_{B}$ . 
Therefore  the Holstein-Primakoff transformation is valid for our analytical results.

On the other hand, for  the limit $N\rightarrow 1$, the maximum power shows a  lower bound of such advantage, see   Fig.\ref{lmgEP}(b). 
The evolution of the system can be easily obtained for $N=1$ case with initial state $\vert m,\uparrow\rangle$.
The energy $\Delta E_{B}=B\sin^2(\sqrt{b_{N_B,m+1}}A t)$ and power $P_{B}= B\sin^2(\sqrt{b_{N_B,m+1}}A t)/t$, so that the maximum of power is  given by $P_{max}\approx 0.72B \sqrt{b_{N_B,m+1}}A$  for the charging time $1.16/\sqrt{b_{N_B,m+1}}A$.
According to the previous setting, the initial state of the battery spins are in lowest state $m\rightarrow 0 $ that gives $\sqrt{b_{N_B,m+1}}=\sqrt{N_B}$ and leads to $P_{max} \propto \sqrt{N_B}$.
This consists with the numerical result given  in Fig.\ref{lmgEP}(b), i.e. the scaling exponent  $\alpha$ varies  from the lower  bound  $\alpha =1/2$  to the upper  bound $3/2$ when the number of charger spins $N$ changes from  small to  the thermodynamic limits, i.e. $N\gg1$ and $N_B\gg1$, while the condition $N\gg N_B$ holds. 



\section{IV. Conclusion}
We have studied numerically and analytically the high-power quantum battery through the multiple central spin model. 
The advantage of quantum battery has been demonstrated through the maximum power of the quantum battery ${P}_{max}=0.72BA\sqrt{N}N_{B}^{\alpha }$ that exhibits a universal power-law dependence of the battery cells (spins) under the condition  $N_B\ll N$. 
Such a power-law relation is analytically derived by the quantum battery of the Tavis-Cummings  type. 
We also have observed that the power-law exponent of the battery power depends on the number of charger spins $N$,
namely the scaling exponent $\alpha$ varies with the bath spin numbers $N$ 
from the lower bound $\alpha =1/2$ to the upper bound $\alpha =3/2$.
 From the maximum power (\ref{single-central-s-P}) of the single central spin battery, we see clearly the maximum power of $N_B$ incoherent  quantum batteries of single central spin systems is given by $P_{ max} \approx 0.72B A \sqrt{N} N_B$.
 Therefore, a quantum advantage is revealed from the  maximum power  Eq. (\ref{pmaxhp}) of the quantum battery of the $N_B$ central spins.
 In the latter case, coherence of the $N_B$ central spins is naturally  created by the interaction between the battery and charger spins. 
In the Appendix, we have presented the analytical results of the quantum battery with $N_B=1$ and an introduction to our  numerical method. 
Our results display the role of how both the charger and battery are capable to enhance the quantum advantage  of the Tavis-Cummings type systems. 
Our rigorous results of dynamical energy transfer shed lights on the design of quantum batteries.

\textit{Acknowledgements.  }W.B.H. acknowledges support from NSAF (Grant No. U1930402). X.W.G. is supported by the  NSFC grant  No.\ 11874393, and the National Key R\&D Program of China  No.\ 2017YFA0304500. S.C. acknowledges support from NSFC (Grants No. 11974040 and No. 1171101295) and the National Key R\&D Program of China No. 2016YFA0301200. H. Q. L. acknowledges financial support from National Science Association Funds U1930402 and NSFC 11734002, as well as computational resources from the Beijing Computational Science Research Center.

\newpage \widetext

\begin{center}
\textbf{\large Appendix }
\end{center}

\setcounter{equation}{0} \setcounter{figure}{0} 
\makeatletter

\renewcommand{\thefigure}{A\arabic{figure}} \renewcommand{\thesection}{SM%
\arabic{section}} \renewcommand{\theequation}{A\arabic{equation}}



\section{The explicit forms of the charging energy}
For the special  case $N_{B}=1$, the Hamiltonian can be written as $2\times2 $ matrix in the basises $ \vert \downarrow \rangle \vert n\rangle,  \vert \uparrow \rangle \vert n-1\rangle$, here $n=1,2,...,N$
\begin{equation}
H_n=\left(  
\begin{array}{cc}
 \frac{B-h}{2}+(n-1-N/2)\Delta &\sqrt{b_{N,n}}A \\
\sqrt{b_{N,n}}A &  -\frac{B-h}{2}-(n-N/2)\Delta \\
\end{array}
\right)
\end{equation}
It is easy to diagonalize above small matrix $H$ analytically to obtain the evolution operator $U(t)=\exp(-iHt)$. The wave function  can be derived by $\vert \psi(t)\rangle=U(t)\vert \Phi_{0}\rangle$.
The Hamiltonian can be written as $H_n=\left( \Delta_{n}/2 \right) \hat{\sigma}_{z}+\sqrt{b_{N,n}}A \hat{\sigma}_{x}+C$, here $C$ is constant. 
The  evolution operator $U(t)$ can be obtained by using property of Pauli matrix namely $\exp(i \theta \hat{n}\cdot \hat{\sigma})=\cos(\theta) I+i \sin(\theta) \hat{n}\cdot \hat{\sigma}$. 
It is  $ U(t)=\cos( \Omega_{n}t/2) I-i \sin( \Omega_{n}t/2)\left[\left( \Delta_{n}/\Omega_{n} \right)\hat{ \sigma}_{z}+2\left( \sqrt{b_{N,n}}A/\Omega_{n}\right) \hat{ \sigma}_{x}\right]$, where 
\begin{eqnarray}
\Delta_{n}&=&B-h+(2n-1-N)\Delta,\nonumber \\
 \Omega_{n}&=&\sqrt{\Delta_{n}^2+4b_{N,n}A^2}.\nonumber
 \end{eqnarray}
 By using  evolution operator $U(t)$ act on the initial state $\vert \downarrow \rangle \vert n \rangle$, we obtain the wave function of the time finally.

We explicitly  rewrite the wave function for $N_B=1$ case Eq.(\ref{wave_function}) as
\begin{equation}
\vert \psi(t)\rangle = e^{-i\theta t}\Big[P_{\uparrow}^{n}(t) \vert \uparrow \rangle \vert n-1\rangle+ P_{\downarrow}^{n} (t) \vert \downarrow \rangle \vert n\rangle \Big].  
\end{equation}
here the global phase $\theta$ can be omitted and two amplitudes are given by
\begin{eqnarray*}\label{amplitude_P}
	P_{\uparrow}^{n}(t)&=&-i\frac{2\sqrt{b_{N,n}}A}{ \Omega_{n}}\sin( \frac{\Omega_{n}t}{2}),\\
	P_{\downarrow}^{n}(t)&=& i\frac{\Delta_{n}}{ \Omega_{n}}\sin( \frac{\Omega_{n}t}{2})+\cos( \frac{\Omega_{n}t}{2}).
\end{eqnarray*}
The wave function satisfies the normalization condition, namely $|P_{\uparrow}^{n}(t)|^2+|P_{\downarrow}^{n}(t)|^2=1$. And the parameters are denoted by $\Delta_{n}=B-h+(2n-1-N)\Delta,\Omega_{n}=\sqrt{\Delta_{n}^2+4b_{N,n}A^2}$.
The density matrix for the system can be obtained as
\begin{eqnarray}
	\rho(t)&=&\left|\Psi(t)\right>\left<\Psi(t)\right|\nonumber\\
	&=&P^n_{\uparrow}(t) {P_{\uparrow}^{n}(t)}^{\ast}\left|\uparrow \right> \left|n-1 \right>\left<n-1 \right| \left<\uparrow  \right| +P^n_{\uparrow}(t) {P_{\downarrow}^{n}(t)}^{\ast}\left|\uparrow \right> \left|n-1 \right>\left<n  \right| \left<\downarrow \right| \nonumber\\
	&&+P^n_{\downarrow}(t) {P_{\uparrow}^{n}(t)}^{\ast}\left|\downarrow \right> \left|n \right>\left<n-1 \right| \left<\uparrow \right| + P^n_{\downarrow}(t) {P_{\downarrow}^{n}(t)}^{\ast}\left|\downarrow \right> \left|n \right>\left<n \right| \left<\downarrow  \right|. 
\end{eqnarray}
Then the reduced density matrices $\rho_B$ and $\rho_C$ are given respectively as 
\begin{eqnarray}
	\rho_B(t) &=& {\tr}_C\Big[\left|\Psi(t)\right>\left<\Psi(t)\right| \Big]\nonumber\\
	&=&P^n_{\uparrow}(t) {P_{\uparrow}^{n}(t)}^{\ast}\left|\uparrow \right> \left< \uparrow \right|+ 
	P^n_{\downarrow}(t) {P_{\downarrow}^{n}(t)}^{\ast} \left|\downarrow \right> \left< \downarrow \right|,\\
	\rho_C(t) &=& {tr}_B\Big[\left|\Psi(t)\right>\left<\Psi(t)\right| \Big]\nonumber\\
	&=& P^n_{\uparrow}(t) {P_{\uparrow}^{n}(t)}^{\ast}\left| n-1 \right> \left< n-1 \right|+ 
	P^n_{\downarrow}(t) {P_{\downarrow}^{n}(t)}^{\ast} \left| n \right> \left< n \right|
\end{eqnarray}
After simple algebra,  we derive the energy of the quantum battery, the energy of charger, and the energy of interaction between charger and battery by substituting the above density matrix into the definition Eq.(\ref{eq_Ebc})
\begin{eqnarray}
E_B(t)&=&\tr[H_B\rho_B(t)]
      =B\Big[\frac{4b_{N,n}A^2}{ \Omega_{n}^2} \sin^{2}( \frac{\Omega_{n}t}{2})-\frac{1}{2}\Big],\label{energy_B}
      \\
E_C(t)&=&\tr[H_C\rho_C(t)]
      =h\Big[(-\frac{N}{2}+n)-\frac{4b_{N,n}A^2}{ \Omega_{n}^2} \sin^{2}( \frac{\Omega_{n}t}{2})\Big],\label{energy_C}
      \\
 E_I(t)&=&\tr[H_I\rho(t)]
       =\Delta(\frac{N}{2}-n)-(B-h)\frac{4b_{N,n}A^2}{ \Omega_{n}^2} \sin^{2}( \frac{\Omega_{n}t}{2}). \label{energy_I}
\end{eqnarray}

\section{The exact diagonalization and fitting   the scaling law}
In this part, we present in  details the exact diagonalization method. According to the action of larger spin operator on the Dicke state, we have 
\begin{eqnarray}
 \mathbf{J^{z}}\vert n \rangle &=& (-\frac{N}{2}+n)\vert n \rangle,\\
\mathbf{J^{-}}  \vert n \rangle &=&  \sqrt{b_{N,n}} \vert n-1 \rangle,\\
\mathbf{J^{+}} \vert n \rangle &=& \sqrt{b_{N,n+1}} \vert n+1 \rangle,
\end{eqnarray}
such that $J^{z},J^{-},J^{+}$ are written as $(N+1)*(N+1)$ matrix, for example, $J^{z}$ and $J^{-}$ are given by 
\begin{equation}
(J^{z})_{mn}= \left\lbrace
\begin{array}{c c c}
  (-\frac{N}{2}+n) ,& for &m=n \\
  0 ,& for & others \\
\end{array}
 \right.
\end{equation}
and 
\begin{equation}
(J^{-})_{mn}= \left\lbrace
\begin{array}{c c c}
   \sqrt{b_{N,n}}  ,& for &m=n-1 \\
  0 ,& for & others \\
\end{array}
 \right. ,
\end{equation}
respectively.
At the same time, the operators $S^{z},S^{-},S^{+}$ can be written as $(N_{B}+1)*(N_{B}+1)$ matrix too. 
By combining the matrix of $J$ and $S$, we obtain the matrix  form of the whole Hamiltonian Eq. \ref{H-B}-\ref{H-I}. 
Thus the dimension of the Hamiltonian in the Dicke basis is $(N_{B}+1)(N+1)$.   
 For $N_{B}\le 40$, and $N\le 300$, the Hamiltonian can be diagonalized directly to obtain the evolving operator $U(dt)=\exp(-iH dt)$ with suitable  time step $dt$. 
 The time dependent wave function  can be obtained numerically $\vert \psi(t) \rangle =U(dt) \cdots U(dt) \vert \Phi_{0} \rangle$. Then according to Eq.(2) and (3), the energy and power can be computed.

\textbf{Scaling relation.} The scaling relation of the maximal power of battery reads
\begin{equation}
P_{max}\propto \beta (N) N_{B}^{\alpha}.
\end{equation}
By taking logarithm, we use linear fitting to obtain the scaling exponent $\alpha$
\begin{equation}
\log(P_{max})= \alpha \log(N_{B})+\log(\beta (N)),
\label{app_plog}
\end{equation}
where $\beta $ is a constant for a fixed $N$. 
 In the numerical fitting in Fig.~(\ref{lmgEP}), we fixed the range of $N_{B}$ in $[1,80]$. Since the total  energy conservation,  the energy reaches to a saturation point for $N_B>N$, see Fig.\ref{lmgEP}(a).  
Therefore we use the data after the kink  to fit the scaling relation  Eq.(\ref{app_plog}) for the region $N<N_B$  in Fig. \ref{lmgEP}(b). 
Similarly, for the region $N\gg N_B$, we  fit the scaling relation  Eq.(\ref{app_plog}) and do find agreement with our analytical relation (\ref{pmaxhp}), see the main text.




\end{document}